\documentclass[a4paper, 12 pt] {article}

\usepackage{amsmath}
\usepackage{amsfonts}
\usepackage{amsthm}
\usepackage{amscd} 

\newtheorem{thm}{Theorem}[section]
\newtheorem{prop}{Proposition}[section]

\usepackage{graphicx}
\usepackage{epsfig}
\usepackage{graphics}

\newcommand\Div{ \,\text{div}\, } 
\newcommand\Curl{\,\text{\bf curl}\,}
\renewcommand\Re{\,\text{Re}\,}
\renewcommand\Im{\text{Im}\,}

\newcommand{\half}{{\textstyle \frac{1}{2}}}
\newcommand{\fourth}{{\textstyle \frac{1}{4}}}
\newcommand{\tfourth}{{\textstyle \frac{3}{4}}}
\newcommand{\tpi}{{\textstyle \frac{1}{2\pi}}}
\newcommand{\fpi}{{\textstyle \frac{1}{4\pi}}}

\newcommand{\Dvec}{{\bf D}}
\newcommand{\nvec}{{\bf n}}
\newcommand{\nvecp}{{\bf n'}}
\newcommand\grad{{\bf \nabla}}

\newcommand\tvec{{\bf t}}

\newcommand\vvec{{\bf v}}

\newcommand\wvec{{\bf w}}
\newcommand\nuvec{\boldsymbol{\nu}}

\newcommand\Chat{{\bf  \hat C}}

\newcommand\jhat{{\boldsymbol{  \hat \jmath}}}
\newcommand\rhat{{\bf \hat r}}
\newcommand\xhat{{\bf \hat x}}
\newcommand\yhat{{\bf \hat y}}
\newcommand\zhat{{\bf \hat z}}

\newcommand\sgn{\text{sgn}\,}

\newcommand{\rvec}{{\bf r}}
\newcommand{\svec}{{\bf s}}

\newcommand{\Acal}{{\cal A}}

\newcommand{\Rcal}{{\cal R}}
\newcommand{\Emin}{{E_{\text{min}}}}

\newcommand\octant{R} 

\newcommand\tbar{{\bar t}}
\newcommand\wbar{{\bar w}}
\newcommand\ebar{{\bar e}}
\newcommand\kbar{{\bar k}}
\newcommand\Omegabar{{\bar \Omega}}

\begin{document}



\title{Elastic energy for
 reflection-symmetric topologies}

\author{A Majumdar$^{\dag\ \ddag}$, 
JM Robbins$^\dag$
\& M Zyskin$^\dag$ 
\thanks{a.majumdar@bristol.ac.uk, j.robbins@bristol.ac.uk, m.zyskin@bristol.ac.uk}\\
School of Mathematics\\ University of Bristol, University Walk, 
Bristol BS8 1TW, UK\\
and\\
Hewlett-Packard Laboratories,\\
 Filton Road, Stoke Gifford, Bristol BS12 6QZ, UK}

\thispagestyle{empty}
\maketitle

\begin{abstract}
  Nematic liquid crystals in a polyhedral domain, a prototype for
  bistable displays, may be described by a unit-vector field subject
  to tangent boundary conditions.  Here we consider the case of a
  rectangular prism.  For configurations with reflection-symmetric
  topologies, we derive a new lower bound for the one-constant elastic
  energy.  For certain topologies, called conformal and anticonformal,
  the lower bound agrees with a previous result.  For the remaining
  topologies, called nonconformal, the new bound is an
  improvement.  For nonconformal topologies we derive an upper bound,
  which differs from the lower bound by a factor depending only on the
  aspect ratios of the prism.  


\end{abstract}


\newpage

\section{Introduction}\label{sec:intro}





Present-day liquid crystal displays (eg twisted nematic) are based on
{\it monostable} cells, wherein, in the absence of external fields,
the orientations of the liquid crystal molecules assume a single
(spatially varying) mean configuration which is effectively
transparent to incident polarised light.  To produce and maintain
optical contrast, voltage pulses, which reorient the molecules, must
be continually applied.  There is considerable interest in developing
{\it bistable} cells, which support two (and possibly more) stable
liquid crystal configurations with contrasting optical properties.  In
bistable cells, power is needed only to switch between the two states.
One mechanism for engendering bistability 
is the cell geometry \cite{jones, newtonspiller, kg2002};
nematic liquid crystals in prototype cells with
polyhedral geometrical features (eg, ridges, or posts) are found to
support multiple configurations.

As a simple model for such systems, we consider the mean local orientation of a
nematic liquid crystal in a polyhedral domain as described by a director
field $\nvec$ subject to suitable boundary conditions.  The situation
we consider, strong azimuthal anchoring, is described by {\it tangent
  boundary conditions}. Tangent boundary conditions require that, on a
face of the domain, $\nvec$ lies tangent to the face, but is otherwise
unconstrained.  This implies that on the edges of the polyhedron, $\nvec$
is parallel to the edges, and therefore is necessarily
discontinuous at the vertices.  We restrict our attention to
director fields which are continuous away from the vertices (ie, as
continuous as possible).  In this case we can unambiguously
assign an orientation to the director (as the domain is simply
connected), and regard $\nvec$ as a unit-vector field.

In \cite{rz2003}, we give a complete topological classification of
continuous tangent unit-vector fields in a convex polyhedron.  An
extension to the nonconvex and periodic cases, along with a general
procedure for
analysing a large class of such classification problems, is given in
\cite{z2005}.  In \cite{mrz2004a} we obtain a lower bound for the
one-constant energy in terms of certain topological invariants, the
trapped areas.  The case of a rectangular prism is considered in
\cite{mrz2004b}, where we also derive an upper bound for the
equilibrium (infimum) energy for a large family of topologies called
reflection-symmetric conformal and anticonformal.  For these
topologies, the ratio of the upper and lower bounds depends only on
the aspect ratios of the prism.  We also show that  topologically
nontrivial behaviour of configurations close to equilibrium may
concentrate near the edges, or may be smoothly distributed, depending
on the aspect ratios.

In this paper we consider again the case of a rectangular prism, and
improve and extend the previous results of \cite{mrz2004b}.
Specifically, we derive a new lower bound for the energy of
reflection-symmetric topologies, expressed in terms of different
invariants, namely the wrapping numbers.  In general, the new lower
bound is an improvement on the previous one.  We also extend the analysis
to all reflection-symmetric topologies, not just conformal
and anticonformal ones.

While liquid crystal applications are a principal motivation for this
work, the problems are also of intrinsic mathematical interest.
Minimizers of the one-constant energy may be regarded as harmonic maps
from a Euclidean polyhedron to the two-sphere $S^2$.  The study of
harmonic maps between Riemannian manifolds is an extensive field, and
connections to problems in liquid crystals are well known
\cite{brecorlie}.  For manifolds with boundary, the regularity of
minimisers for the Dirichlet problem for harmonic maps with
sufficiently smooth ($C^2$) boundary and Dirichlet data are
investigated in \cite{su}.  However, less appears to be known about
the case of manifolds with Lipschitz boundary, eg domains with
corners, and for natural, eg tangent boundary conditions.
There are recent strong results on the existence,
uniqueness and regularity of minimisers for the Dirichlet problem for harmonic
maps of fixed homotopy type between Riemannian polyhedra for target
spaces of negative curvature \cite{eellsfug}.  However, it appears to
be much more difficult to obtain corresponding results for target
spaces of positive curvature, eg $S^2$, which we encounter in liquid
crystals problems.

The paper is organised as follows.  The topological classification of
tangent unit-vector fields in a rectangular prism is reviewed in
Section~\ref{sec: topology}. We introduce the reflection-symmetric
topologies, which are characterised by certain invariants -- the edge
signs $e$, kink numbers $k$ and trapped area $\Omega$ -- associated
with one of the prism vertices. In Section~\ref{sec: lower bound} we
derive a lower bound for the one-constant elastic energy.  This turns
out to depend on the absolute values of the wrapping numbers (which
may be expressed in terms of $e$, $k$ and $\Omega$).  For certain
topologies, called conformal and anticonformal, for which the wrapping
numbers all have the same sign, the lower bound can be expressed in
terms of the trapped area alone, and coincides with the result
previously derived in \cite{mrz2004a,mrz2004b}.
Conformal and anticonformal topologies are characterised in
Section~\ref{sec: conformal}, where it is shown that these are
precisely the topologies which have conformal and anticonformal
representatives of the type considered in \cite{mrz2004b}.  In Section
\ref{sec: upper bound} we introduce representative configurations for
nonconformal topologies, and derive from them an upper bound for the
elastic energy.  This differs from the lower bound of
Section~\ref{sec: topology} by a factor depending only on the aspect
ratios.  Appendix~\ref{sec:appendix} contains a derivation of a formula for the kink
numbers of conformal and anticonformal configurations.

\section{Reflection-symmetric topologies}\label{sec: topology}


Let us briefly recall the results concerning the classification of
continuous tangent unit-vector fields $\nvec$ on a 
rectangular prism $P$.
For convenience, we let $P$ denote the prism
without its vertices, so that $\nvec$ is continuous on $P$.
For definiteness, we take the prism to be given by
$0 \le r_j \le L_j$, with edge lengths $L_j$
ordered so that
$L_x \ge L_y \ge L_z$.  At each vertex of $P$, denoted $\vvec =
(v_x,v_y,v_z)$, we associate to $\nvec$ a set of topological
invariants, namely the {\it edge signs}, {\it kink numbers} and {\it
  trapped area}.  The edge signs, denoted $e^\vvec_j$,  determine the
signs of $\nvec$ on the edges at $\vvec$ relative to the coordinate
unit vectors, ie
\begin{equation}
  \label{eq:edge sign}
  \nvec(x,v_y,v_z) = e_x^\vvec\, \xhat, \ \ 0 \le x \le  L_x,
\end{equation}
and similarly for $e_y^\vvec$ and $e_z^\wvec$.  (Of course, this designation
is redundant; the edge signs $e^\vvec_j$ and $e^{\wvec}_j$ at vertices
$\vvec$ and $\wvec$ joined by an edge parallel to the $j$-axis are
necessarily the same.)

The integer-valued kink numbers,
denoted $k^\vvec_j$, count the windings of $\nvec$
along a path about $\vvec$ on the face normal to $\jhat$.  
By convention, the paths are taken to be positively oriented with
respect to the outward normal through the centre of the face.  The
minimum possible winding (a net rotation of $\pm \pi/2$) 
is assigned a kink number of zero.  Nonzero windings are
designated positive or negative according to their orientation with respect to
the outward normal (either $\jhat$ or $-\jhat$).  The continuity of $\nvec$ (in particular, the absence of
singularities on the surface of $P$) implies that the kink numbers on
each face satisfy a sum rule \cite{rz2003}; for example, on one of the faces $F$ normal
to $\zhat$,
it turns out that
\begin{equation}
  \label{eq:kinksumrule}
  {\sum_{\vvec\in F}} \left(k^\vvec_z -
  \fourth(-1)^{v_x/L_x} (-1)^{v_y/L_y} e_x^\vvec e_y^\vvec \right) = 0.
\end{equation}
 Analogous rules hold
for the other faces.

The last invariant, the trapped area, denoted $\Omega^\vvec$, is the
oriented area on the unit two-sphere $S^2$ of the image, $\nvec(C^\vvec)$,
of a surface, $C^\vvec$, which separates $\vvec$ from the other
vertices.  That is, letting $(\theta,\phi)$ denote polar coordinates
on $S^2$, 
\begin{equation}
  \label{eq:trapped_area_1}
  \Omega^\vvec = \int_{\nvec(C^\vvec)} \sin\theta \, d\theta \wedge d\phi.
\end{equation}
Expressed as an integral over $C^\vvec$ itself, $\Omega^\vvec$ is given by
\begin{equation}
  \label{eq:trapped_area}
  \Omega^\vvec = \int_{C^\vvec} \Dvec \cdot \Chat^\vvec\, dS.
\end{equation}
Here $\Chat^\vvec$ is the outward-oriented unit normal on $C^\vvec$
($\Chat^\vvec$ points towards $\vvec$) and $dS$ is the area element,
while the vector field $\Dvec(\rvec)$ is given by
\begin{equation}
  \label{eq:D}
D_j = \half 
\epsilon_{jkl} (\partial_k \nvec \times \partial_l \nvec)\cdot \nvec.
\end{equation}
That
(\ref{eq:trapped_area_1}) and (\ref{eq:D}) are equivalent follows from
the fact that
\begin{equation}
  \label{eq:D_and_jacobian}
  \Dvec\cdot\Chat^\vvec = \det d\nvec_{C^\vvec},
\end{equation}
where $d\nvec_{C^\vvec}$ denotes the Jacobian of the restricted map 
$\nvec_{C^\vvec}: C^\vvec \rightarrow S^2$ 
(and, as above, $C^\vvec$ is oriented with
respect to the outward normal).

For a rectangular
prism, the trapped areas are necessarily odd multiples of $\pi/2$
(the
area of a right spherical triangle), and for given values of the edge
signs and kink numbers, the allowed values of the trapped areas differ by
multiples of $4\pi$ (whole coverings of the sphere) -- see
(\ref{eq:trapped_area_and_wrapping_number}) below.  The continuity
of $\nvec$ (the absence of singularities inside $P$) implies the sum
rule
\begin{equation}
  \label{eq:trapped_area_sum_rule}
  \sum_{\vvec} \Omega^\vvec = 0.
\end{equation}

One can show (\cite{rz2003}) that the edge signs, kink numbers and
trapped areas are indeed topological invariants (ie, they are invariant under
continuous deformations of $\nvec$ that preserve the tangent boundary
conditions)
and that two tangent unit-vector fields on $P$ are
homotopic if and only if their invariants are the same.  For
convenience we have slightly adapted the notation of \cite{rz2003}
to suit the case of prisms (the conventions, however, are the same).

In what follows we restrict our attention to a subset of the allowed
prism topologies which we call {\it reflection symmetric}.  Let
$\vvec$ and $\wvec$ denote a pair of vertices related by a reflection
though a midplane of the prism (and therefore joined by an edge).  For
reflection-symmetric topologies, the edge signs at $\vvec$ and $\wvec$
are the same while the kink numbers and trapped areas differ by a
sign.  That is,
\begin{equation}
    \label{eq:reflection-symmetric-def}
    e^{\vvec}_j = e^{\wvec}_j, \quad
    k^{\vvec}_j = -k^{\wvec}_j, \quad
    \Omega^{\vvec} = -\Omega^{\wvec}.
\end{equation}
It follows that at vertices related by two reflections (ie, at
diagonally opposite corners of a face), the invariants are the same,
while the invariants of vertices related by three reflections (at
diagonally opposite corners of the prism) are related as in
(\ref{eq:reflection-symmetric-def}).

For reflection-symmetric topologies the invariants are determined by
their values at a single vertex.  For definiteness we take this vertex
to be the origin, and henceforth denote the invariants simply as
$(e,k,\Omega)$.  The surface separating the origin from the other
vertices 
will be denoted by $C$.  It is
straightforward to check that (\ref{eq:reflection-symmetric-def})
implies that the sum rules (\ref{eq:kinksumrule}) and
(\ref{eq:trapped_area_sum_rule}) are automatically satisfied.

The terminology stems from the fact that every reflection-symmetric
topology has a reflection-symmetric representative, ie a configuration
$\nvec$ which is symmetric under reflections through the
mid planes,
\begin{equation}
  \label{eq:reflection-symmmetric_n}
  \nvec(x,y,z) = \nvec(L_x-x,y,z) = \nvec(x,L_y-y,z) = \nvec(x,y,L_z-z).
\end{equation}
Let
\begin{equation}
  \label{eq:R}
  R = \{\rvec \,|\, 0 \le r_j \le \half L_j  \}
\end{equation}
denote the octant of the prism with the origin as vertex.  Then a
reflection-symmetric configuration is determined by its values in $R$.
It is straightforward to verify that
(\ref{eq:reflection-symmmetric_n}) implies the relations
(\ref{eq:reflection-symmetric-def}). 
Conversely, given a configuration $\nvecp$ with reflection-symmetric
topology $(e,k,\Omega)$ 
but which is not itself reflection symmetric, we can construct a
reflection-symmetric configuration $\nvec$ satisfying
(\ref{eq:reflection-symmmetric_n}) with invariants $(e,k,\Omega)$
(just take $\nvec = \nvecp$ in the prism octant $R$
and define $\nvec$ elsewhere via (\ref{eq:reflection-symmmetric_n})).

In \cite{rz2003} we introduced certain additional integer-valued
topological invariants, called {\it wrapping numbers}.  As the
preceding discussion implies, the wrapping numbers are not independent
of the edge signs, kink numbers and trapped areas, but rather can be
expressed in terms of them.  We briefly recall the definition and
properties of the wrapping numbers, as they are central to the
discussion to follow.

Let $\svec$ denote a regular value of $\nvec$ restricted to $C$.  That
is, on $C$, there is a finite number of points where $\nvec$ takes the
value $\svec$, and, at any such point, the Jacobian of the map $\nvec_C:
C \rightarrow S^2$ is nonsingular, so that, from
(\ref{eq:D_and_jacobian}), $(\Dvec\cdot \Chat)(\svec) \ne 0$.  The
wrapping number at $\svec$, denoted $w(\svec)$, is a signed count of
the pre-images of $\svec$ on $C$, denoted $\rvec_p$, the sign  determined by
whether $\nvec_C$ is orientation-preserving $(+)$
or reversing $(-)$ at $\rvec_p$.  Thus
\begin{equation}
  \label{eq:wrapping_number_defn}
  w(\svec) = 
\sum_{p} \sgn
  [(\Dvec\cdot \Chat)(\rvec_p)].
\end{equation}
To express the wrapping number in terms of the other invariants
\cite{rz2003, mrzcor}, let $U_{p\epsilon}$ denote the disk of radius
$\epsilon$ about $\rvec_p$ on $C$, and let $C - \sum_p U_{p\epsilon}$
denote $C$ with these disks excised.  Let $\partial C$ denote the
boundary of $C$ and $\partial U_{p\epsilon}$ the boundary of  $U_{p\epsilon}$.
Choosing $\svec$ as the south pole of the polar
angles $(\theta,\phi)$  in
(\ref{eq:trapped_area_1}) and using Stokes' theorem, we get that
\begin{equation}
  \label{eq:wrap_calculation_1}
  \Omega = \lim_{\epsilon\rightarrow 0} 
\int_{\nvec\left(C - \sum_p U_{p\epsilon}\right)} \sin\theta \, d\theta
\wedge d\phi
= \left(
\int_{\nvec\left(\partial C\right)} - \lim_{\epsilon\rightarrow 0}  
\int_{\nvec\left(\sum_p\partial U_{p\epsilon}\right)}\right)
(1 - \cos\theta)d\phi.
\end{equation}
$\nvec(\sum_p \partial U_{p\epsilon})$ consists of
$p$ small circuits about $\svec$, and the integral over these circuits
in (\ref{eq:wrap_calculation_1}) 
gives, in the limit $\epsilon \rightarrow 0$, $4\pi$ times
 $w(\svec)$.  
$\nvec(\partial C)$ consists of
the spherical right triangle with vertices $e_j
\jhat$, $j = x, y, z$, along with  
$k_j$ circuits of the great circle normal to
$\jhat$.
Each great circle contributes $\pm 2\pi$ to the integral in
(\ref{eq:wrap_calculation_1}) according to its orientation with
respect to the polar axis through $\svec$, while the spherical
triangle contributes $\pm \pi/2$ (its signed area) or $\pm 7\pi/2$
according to whether or not it encloses $\svec$.  
Keeping track of signs one gets
\begin{equation}
   \label{eq:trapped_area_and_wrapping_number}
 w(\svec) = \fpi \Omega + \half \sum_j \sigma_j k_j + e_x e_y e_z \times
  \begin{cases}
    -\frac78,& \text{if}\ \sigma_j = \sgn e_j\ \text{for all $j$},\\
    +\frac18, &\text{otherwise}.
  \end{cases}
\end{equation}
where $\sigma_j = \sgn s_j$.  From
(\ref{eq:trapped_area_and_wrapping_number}) it is clear that
$w(\svec)$ is a topological invariant.
In fact, $w(\svec)$ can be defined so long as $\svec$ does not lie in a
coordinate plane (ie, even if $\svec$ is not a regular value of
$\nvec$) as the degree of a certain continuous $S^2 \rightarrow S^2$ map
constructed by gluing $\nvec: C\rightarrow S^2$ to a reference
map which coincides with $\nvec$ on the boundary $\partial C$
\cite{rz2003}.

(\ref{eq:trapped_area_and_wrapping_number}) also implies that
$w(\svec)$ depends
only on the signs of the components of $\svec$, ie on the
octant of $S^2$ to which $\svec$ belongs.  In what follows, we label
octants by
a triple of signs $\sigma = (\sigma_x,\sigma_y,\sigma_z)$, so that
$O_\sigma$ denotes the octant $\{\svec \,|\, \sgn s_j = \sigma_j\}$.
For
convenience we let
$w_\sigma$ denote the value of $w(\svec)$ for $\svec \in O_\sigma$.
Summing over octants in (\ref{eq:trapped_area_and_wrapping_number}),
we get that
\begin{equation}
  \label{eq:sum_of_ws}
  \Omega = \half \pi \sum_\sigma w_\sigma
\end{equation}
(the terms in
(\ref{eq:trapped_area_and_wrapping_number}) involving $e_j$ and $k_j$ 
cancel in the sum). 



\section{Lower bound for the elastic energy}\label{sec: lower bound}
In the continuum theory of nematic liquid crystals \cite{degennes}, the elastic, or
Frank-Oseen, energy of a
configuration $\nvec$
is given by 
\begin{multline}
  \label{eq:elastic}
  E(\nvec) = \int_P \big [ K_1 (\Div \nvec)^2 + K_2 (\nvec
  \cdot \Curl \nvec)^2 +   K_3
  (\nvec\times \Curl \nvec)^2\\ +  K_4 \Div ((\nvec\cdot
  \grad)\nvec - (\Div \nvec) \nvec)\big ]\, dV.
\end{multline}
Tangent boundary conditions 
imply that the contribution from the $K_4$-term, which is a pure
divergence, vanishes.  In the so-called one-constant approximation, 
the remaining elastic constants $K_1$, $K_2$ and $K_3$ are taken to be
the same.  In this case, (\ref{eq:elastic}) simplifies to
\begin{equation}
  \label{eq:one-constant}
  E(\nvec) = \int_P  (\grad \nvec)^2\,  dV = 
K \int_P  \sum_{j=1}^3 (\partial_j \nvec)^2  \, dV.
\end{equation}
We shall use the one-constant approximation in what follows.
%

Let $\Emin(e,k,\Omega)$ denote the minimum (infimum) energy for
configurations with reflection-symmetric topology $(e,k,\Omega)$.
In  \cite{mrz2004b} we obtained the lower bound
  \begin{equation}
  \label{eq:old_bound}
  \Emin(e,k,\Omega) \ge 8 L_z |\Omega|.
\end{equation}
In view of (\ref{eq:sum_of_ws}), this may be written as
  \begin{equation}
  \label{eq:old_bound2}
  \Emin(e,k,\Omega) \ge 4\pi L_z \left|\sum_\sigma w_\sigma\right|.
\end{equation}
Here we derive a new lower bound which, in general, is
an improvement on (\ref{eq:old_bound}).
\begin{thm}\label{thm: lower bound}
 \begin{equation}
  \label{eq:infimum_result}
  \Emin(e,k,\Omega) \ge  4 \pi  L_z \sum_{\sigma} |w_\sigma|.
\end{equation}
\end{thm}
\begin{proof}
  Let $\nvec$ be a configuration with reflection-symmetric topology
  $(e,k,\Omega)$ for which the  energy
  (\ref{eq:one-constant}) is finite.  As shown in
  \cite{mrz2004a}, we can, without loss of generality, take
  $\nvec$ to be smooth (smooth configurations are dense in the space
  of finite-energy configurations with respect to the energy norm).

We can assume that the energy of $\nvec$ in
$R$ is not more than its energy in any other octant of the prism
(we can replace $\nvec(\rvec)$ by $\nvec(\Rcal\cdot \rvec)$ for a
product $\Rcal$ of reflections through mid planes; the reflected configurations
have
the same topology and energy as $\nvec$).
Then
\begin{equation}\label{eq:eq}
  E(\nvec) \ge 8 \int_R  |\grad \nvec|^2 \, dV \ge 8 \int_{r \le
  L_z/2}  |\grad \nvec|^2 \, dV,
\end{equation}
where the last integral is taken over the positive octant  of the ball
of radius $ L_z/2$ about the origin.
Using the local inequality for the
energy density \cite{brecorlie, mrz2004a, mrz2004b}, 
\begin{equation}
  \label{eq:grad n^2}
  (\grad\nvec)^2 \ge  2|\Dvec| \ge 2|\Dvec\cdot \rhat|, 
\end{equation}
we get that
  \begin{equation}
    \label{eq:energy_2}
    E(\nvec) \ge 16 \int_{r \le L_z/2} |\Dvec(\rvec)\cdot\rhat| \, dV = 
16\int_0^{L_z/2} dr \int_{\rvec\in C_r}|\Dvec(\rvec)\cdot\rhat| \,  dS_r .
  \end{equation}
Here, $C_r$ is the positive octant of the sphere of radius $r$ about
the origin, with
area element $dS_r$.

We partition $C_r$ into preimages of the octants
$O_\sigma$ of $S^2$, writing
\begin{equation}
    \label{eq:energy_2a}
    E(\nvec) 
\ge 16\int_0^{L_z/2} dr \int_{\rvec\in C_r}\left(\sum_\sigma
\int_{O_\sigma} {\bf ds} \, \delta_{S^2}(\nvec(\rvec),\svec)\right)
|\Dvec(\rvec)\cdot\rhat|\,  dS_r.
  \end{equation}
  Here $\delta_{S^2}$ is the normalised Dirac delta-function on $S^2$,
  so that $\int_{O_\sigma} {\bf ds} \,
  \delta_{S^2}(\nvec(\rvec),\svec)$ equals one if $\nvec(\rvec) \in
  O_\sigma$ and is zero otherwise.  We interchange the integrals over
  $\rvec \in C_r$ and $\svec$ and take the absolute value outside
  these integrals to obtain
\begin{equation}
  \label{eq:eq:energy_2b}
E(\nvec) 
\ge 16\int_0^{L_z/2} dr
\sum_\sigma 
 \left| \int_{O_\sigma} 
 {\bf ds} \int_{\rvec\in C_r}  \delta_{S^2}(\nvec(\rvec),\svec)
\Dvec(\rvec)\cdot\rhat \, dS_r\,\right| .
\end{equation}
For $\svec$ a regular value of $\nvec$ (by Sard's theorem,
regular values are of full measure), we get that
\begin{equation}
  \label{eq:delta_integral}
  \int_{\rvec\in C_r}  \delta_{S^2}(\nvec(\rvec),\svec)
\, dS_r = \sum_{p} \left|\det d\nvec_{C_r}(\rvec_p)\right|^{-1} = 
\sum_{p} \left|\Dvec(\rvec_p)\cdot\rhat\right|^{-1},
\end{equation}
where the sum is taken over the preimages $\rvec_p \in C_r$ of $\svec$,
and we have used (\ref{eq:D_and_jacobian}).  Substituting into
(\ref{eq:eq:energy_2b}), we get that
\begin{equation}
  \label{eq:energy_2c}
  E(\nvec) \ge 16\int_0^{L_z/2} dr \sum_\sigma  \left| \int_{O_\sigma} 
 {\bf ds} \sum_{p}  \sgn (\Dvec(\rvec_p)\cdot \rhat)
\right|.
\end{equation}
From (\ref{eq:wrapping_number_defn}), the sum over $p$ is just 
the wrapping number $w(\svec) =
w_\sigma$, so that the
integral over $\svec$ trivially gives a factor of $\pi/2$ (the area of
$O_\sigma$). Then the integral over $r$ trivially gives a factor of
$ L_z/2$.
The required result (\ref{eq:infimum_result}) follows.

\end{proof}


\section{Conformal and anticonformal topologies}\label{sec: conformal}

The new bound (\ref{eq:infimum_result}) agrees with the previous bound
(\ref{eq:old_bound2}) for topologies where $\sum_\sigma |w_\sigma| =
\left |\sum_\sigma w_\sigma\right|$, ie where the nonzero wrapping
numbers all have the same sign.  We will say that a
reflection-symmetric topology is {\it conformal} if $w_\sigma \le 0$
for all $\sigma$, {\it anticonformal} if $w_\sigma \ge 0$ for all
$\sigma$, and {\it nonconformal} if neither of these conditions holds.
Thus, the new bound constitutes an improvement for nonconformal
topologies.

It is useful to characterise the conformal and anticonformal topologies
directly in terms of the invariants $(e,k,\Omega)$.  
\begin{prop}
\label{prop: 2.1}
Define functions $\Omega_\chi(e,k)$, where $\chi =
\pm$, as follows:\\

\noindent For $\chi e_x e_y e_z =
  1$,
  \begin{equation}
    \label{eq:Omega_+}
    \Omega_\chi(e,k) = 2\pi \sum_j |k_j| + 2\pi
    \begin{cases}
     +\frac74,& \text{if}\ \chi e_j  k_j \le 0\ \text{for all}\ j,\\
     -\frac14,& \text{otherwise}.
    \end{cases}
  \end{equation}
For $\chi e_x e_y e_z = -1$, 
 \begin{equation}
    \label{eq:Omega_-}
    \Omega_\chi(e,k) = 2\pi \sum_j |k_j| - 2\pi
    \begin{cases}
     +\frac74,& \text{if}\ \chi e_j  k_j < 0\ \text{for all}\ j,\\
     -\frac14,& \text{otherwise}.
    \end{cases}
  \end{equation}
  Then the reflection-symmetric topology $(e,k,\Omega)$ is conformal
  if and only if $\Omega \le -\Omega_-(e,k) $ and anticonformal if and
  only if $\Omega \ge \Omega_+(e,k)$.  If equality obtains, ie $\Omega = -\Omega_-(e,k)$ or
  $\Omega = \Omega_+(e,k)$, then at least one wrapping number must vanish.
\end{prop}
\begin{proof}
The condition $\chi w_\sigma \ge 0$ for all $\sigma$ is equivalent to 
$(e,k,\Omega)$ being conformal ($\chi = -$) or anticonformal ($\chi = +)$.
From (\ref{eq:trapped_area_and_wrapping_number}),
  $\chi w_\sigma \ge 0$ for all $\sigma$ if and only if $\chi\Omega \ge
  \Omega_\chi(e,k)$, where
  \begin{equation}
    \label{eq:Omega_I_def}
    \Omega_\chi(e,k) = 2\pi \max_{\sigma} \left( -\chi \sum_j \sigma_j k_j +
    \chi e_x e_y e_z \times
      \begin{cases}
        +\frac74,& \text{if $\sigma_j = e_j$ for all $j$} \\
          -\frac14,& \text{otherwise}
      \end{cases}\right),
  \end{equation}
  with $\chi \Omega = \Omega_\chi(e,k)$ if and only if $w_\sigma = 0$
  for some $\sigma$.  In (\ref{eq:Omega_I_def}), to realise the
  maximum we may take, for all $j$ such that $k_j \ne 0$, $\sigma_j =
  -\chi \sgn k_j$, and thereby replace $-\chi \sigma_j k_j$ by $|k_j|$
  for all $j$.  Thus,
 \begin{equation}
    \label{eq:Omega_I_2}
    \Omega_\chi(e,k) = 2\pi \sum_j |k_j| +  \max_{\sigma_j \,|\, k_j = 0}
    \chi e_x e_y e_z \times
      \begin{cases}
        +\frac72\pi,&\text{if $\sigma_j = e_j$ for all $j$},\\
          -\frac12\pi,& \text{otherwise}
      \end{cases}.
  \end{equation}
  
  It remains to maximise the second term in (\ref{eq:Omega_I_2}) with
  respect to the $\sigma_j$'s for which $k_j = 0$.  Suppose that $\chi
  e_x e_y e_z = 1$.  Then, provided $\sigma_j = e_j$ for all $k_j \ne
  0$, ie provided $-\chi \sgn k_j = e_j$ for all $k_j \ne 0$, the
  maximum value attained by the second term in (\ref{eq:Omega_I_2}) is
  $7\pi/2$.  Otherwise,
the maximum is $-\pi/2$.  This is in
  accord with (\ref{eq:Omega_+}).  Next, suppose that $\chi e_x e_y e_z
  = -1$.  The
  maximum value attained by the second term in (\ref{eq:Omega_I_2}) is 
$\pi/2$ unless all the $k_j$'s are
  nonzero and $-\chi \sgn k_j = e_j$, in
  which case the maximum is $-7\pi/2$.
  This is in accord with (\ref{eq:Omega_-}).
\end{proof}



In \cite{mrz2004b} we introduced certain reflection-symmetric
configurations in $P$ which we called conformal and anticonformal.
We show next that the conformal and anticonformal
topologies are precisely those which have conformal and anticonformal
representatives.
To proceed, we  briefly recall the properties of conformal configurations
(\cite{mrz2004a}) (as discussed below, the treatment of 
anticonformal configurations is analogous).
A reflection-symmetric configuration $\nvec$ is said to be {\it conformal} if, 
in the prism octant $R$,
i) $\nvec$ is radially constant, ie $\nvec(\lambda
\rvec) = \nvec(\rvec)$, and ii) $\nvec$ is conformal, ie the map
$\tvec\mapsto \grad_t \nvec(\rvec)$ from vectors $\tvec$ orthogonal to
$\rhat$ to vectors $\grad_t \nvec(\rvec)$ orthogonal to $\nvec(\rvec)$
preserves orientation, angles and ratios of lengths (or else
vanishes).  

Conformal configurations 
are conveniently represented
via stereographic projection as analytic functions $f(w)$,
\begin{equation}
  \label{eq:stereograph}
  \left(\frac{n_x + in_y}{1 + n_z}
  \right)(x,y,z) = 
f\left(
\frac{x + iy}{r + z}
\right).
\end{equation} 
The domain of $f(w)$ is the quarter-unit-disk $Q$ given by $|w| \le
1$, $0 \le \Re w \le 1$ and $0 \le \Im w \le 1$.  The boundary of
$Q$ consists of the real
interval
$0 \le w \le 1$ (which corresponds to the $xz$-face of $R$), the imaginary interval
$0 \le -iw \le 1$ (which corresponds the $yz$-face), and the circular arc $|w|
= 1$, where $0 \le \arg w \le \pi/2$ (which corresponds to the $xy$-face).
Tangent boundary conditions imply that i) $f(w)$ is real for $w$ real,
 ii) $f(w)$ is imaginary for $w$ imaginary, 
and iii) $|f(w)| = 1$ if $|w| = 1$.
Assuming that $f(w)$ has a meromorphic extension to the extended complex
plane, these conditions imply that
if $w_*$ is a zero of $f$, then 
$-w_*$ and $\wbar_*$ are zeros, while $1/\wbar_*$ is a pole. 
The meromorphic functions which satisfy these conditions are rational
functions of the following form:
\begin{multline}
    \label{eq:w(z)}
    f(w) = \epsilon w^n 
  \prod_{j=1}^a\left(
  \frac{w^2 - r_j^2}{r_j^2 w^2 - 1}
  \right)^{\rho_j}
  \prod_{k=1}^b\left(
  \frac{w^2 + s_k^2}{s_k^2 w^2 + 1}
  \right)^{\sigma_k}\times\\
\times  \prod_{l=1}^c\left(
  \frac{(w^2 - t_l^2)(w^2 -  \tbar_l^2)}
  {(t^2_l w^2 - 1)({\tbar}_l^2w^2  - 1)}
  \right)^{\tau_l}.
\end{multline}
Here, $\epsilon = \pm 1$ and $n$, an odd integer, gives the order of
the zero or pole of $f$ at the origin.  $a$ is the number of zeros of $f$
($\rho_j = 1$) and poles of $f$ ($\rho_j = -1$) on the real interval $(0,1)$,
with positions $r_j$ ordered so that $0 < r_1 \le \cdots \le r_a < 1$.
Similarly, $b$ is the number of zeros of $f$ ($\sigma_k = 1$) and poles
of $f$ ($\sigma_k = -1$) on the imaginary interval $(0,i)$, with positions
$is_k$ ordered so that $0 < s_1 \le \cdots \le s_b < 1$.  Finally, $c$
is  the number of zeros of $f$ ($\tau_l = 1$) and poles of $f$
($\tau_l = -1$) in the interior of $Q$, with positions $t_l$. 

The edge signs, kink numbers and trapped area of
conformal configurations are given by
\begin{equation}
  \label{eq:edge signs}
e_x = \epsilon(-1)^a,\
e_y = \epsilon(-1)^b(-1)^{(n-1)/2},\ 
e_z = \sgn\, n,
\end{equation}
\begin{align}
  \label{eq:ks}
  k_x &= -\half (-1)^{b} e_y 
\left(
{ \sum_{k = 1}^b} (-1)^{k}\sigma_k + 
\half (1 - (-1)^{b}) e_z\right)
,\nonumber\\
  k_y &= -\half (-1)^a e_x 
\left(
{ \sum_{j = 1}^a} (-1)^{j} \rho_j + \half (1-(-1)^{a})e_z
\right)
,\\
k_z &= \fourth\left(e_xe_y - n\right) 
-\half{ \sum_{j=1}^a} \rho_j
-\half{ \sum_{k=1}^b} \sigma_k
-{ \sum_{l=1}^c} \tau_l,\nonumber
\end{align}
and
\begin{equation}
  \label{eq:Omega formula}
  \Omega = -\half (|n| + 2(a + b) + 4c)\pi.
\end{equation}
As explained in \cite{mrz2004b}, the expressions 
for the edge signs and 
the
trapped area are easily derived (in particular, (\ref{eq:Omega
  formula}) follows from consideration of the degree of $f$ on the extended
complex plane).  A derivation of the expression for the
kink numbers, which was deferred in \cite{mrz2004b}, is given here in
Appendix~\ref{sec:appendix}.

Clearly, a conformal configuration has a conformal topology; the
orientation-reversing property ensures that the wrapping numbers
cannot be positive.  Below we establish the converse fact;  every
conformal topology  $(e,k,\Omega)$ has a conformal representative.  
The demonstration splits into four cases according
to the sign of $e_x e_y e_z$ and of $e_j
k_j$.  In each case we exhibit the parameter values, expressed in terms of
$e$, $k$ and $\Omega$, for a particular
conformal configuration.  It is then straightforward to verify -- we
omit the explicit demonstration -- that the specified parameters are
admissable (ie, that $n$ is an odd integer; $a$, $b$, $c$ are
nonnegative integers; $\epsilon$, $\rho_j$, $\sigma_k$, $\tau_l$ are
signs), and that, with these parameters, the values of the invariants
given by (\ref{eq:edge signs})--~(\ref{eq:Omega formula}) are just
$(e,k,\Omega)$.  Deriving the exhibited values involves a systematic
and slightly tedious investigation of (\ref{eq:edge
  signs})--~(\ref{eq:Omega formula}).  For the sake of brevity, these
details are also omitted.
\\

\noindent {\it Case 1a. $e_x e_y e_z  = 1$ and $e_j k_j > 0$ for all
  $j$.} Let
\begin{gather}
  \epsilon = -e_x,\quad
  n = e_z,\nonumber\\
  a = 2|k_y| - 1, \quad \rho_j = (-1)^j e_z,\nonumber\\
  b = 2|k_x| - 1, \quad \sigma_k = (-1)^k e_z, \nonumber\\
  c = -\tpi \Omega - |k_x| - |k_y| + \tfourth, \quad \tau_l = 
  \begin{cases}
    -e_z,& l < |k_z|,\\
    (-1)^l,& l \ge |k_z|.
  \end{cases}
\end{gather}
Note that (\ref{eq:trapped_area_and_wrapping_number})
implies that $c - (|k_z| - 1)$ is nonnegative and even.  Here and in the cases to follow, we
do not specify the positions of the zeros and poles explicitly.\\

\noindent {\it Case 1b. $e_x e_y e_z  = 1$ and $e_j k_j \le 0$ for some $j$.} 
Without loss of generality, we may assume that $j = z$.  
This follows from considering the fractional linear transformation
\begin{equation}
  \label{eq:frac_rot}
  r(w) = \frac{i-w}{i+w}
\end{equation}
which maps $Q$ onto itself while cyclically permuting its vertices
($r$ corresponds to the $2\pi/3$-rotation on $S^2$ about the axis (1,1,1)).
Therefore, if $f$ is a conformal configuration, so
is $\tilde f$ given by
\begin{equation}
  \label{eq:transformed}
  \tilde f = r \circ f \circ r^{-1}.
\end{equation}
It is easily verified that $\tilde f$ and $f$ have the same 
trapped areas
while their edge signs and kink numbers are related by cyclic
permutation,
\begin{equation}
  \label{eq:permute}
  \tilde e = (e_z, e_x, e_y), \quad  \tilde k = (k_z, k_x, k_y).
\end{equation}

Letting $e_z k_z\le 0$, we take
\begin{gather}
  \epsilon = e_x,\quad
  n = (4 |k_z| + 1)e_z,\nonumber\\
  a = 2|k_y|, \quad \rho_j = -(-1)^j e_x \sgn k_y,\nonumber\\
  b = 2|k_x|, \quad \sigma_k = -(-1)^k e_y \sgn k_x,\nonumber\\
  c = -\tpi\Omega - |k_x| - |k_y| - |k_z| - \fourth, \quad \tau_l = (-1)^l.
\end{gather}
Note that (\ref{eq:trapped_area_and_wrapping_number}) implies that $c$
is nonnegative and even.
\\

\noindent {\it Case 2a. $e_x e_y e_z = -1$ and $e_j k_j < 0$ for some $j$.}
As in Case 1b, without loss of generality, we may take $e_z k_z < 0$.  Let
\begin{gather}
  \epsilon = e_x,\quad
  n = -(4 k_z + e_z),\nonumber\\
  a = 2|k_y|, \quad \rho_j = -(-1)^j e_x \sgn k_y,\nonumber\\
  b = 2|k_x|, \quad \sigma_k = -(-1)^k e_y \sgn k_x,\nonumber\\
  c = -\tpi\Omega - |k_x| - |k_y|-|k_z| + \fourth, \quad \tau_l = (-1)^l.
\end{gather}
Note that (\ref{eq:trapped_area_and_wrapping_number}) implies that $c$
is nonnegative and even.\\

\noindent {\it Case 2b. $e_x e_y e_z = -1$ and $e_j k_j \ge 0$ for all
  $j$.} Let
\begin{gather}
  \epsilon = e_x,\quad n =3 e_z,\nonumber\\
  a = 2|k_y|,\quad \rho_j = e_z(-1)^j,\nonumber\\
  b = 2|k_x|,\quad \sigma_k = e_z(-1)^k,\nonumber\\
  c = -\tpi \Omega - |k_x| - |k_y| - \tfourth, \quad \tau_l = 
  \begin{cases}
    -e_z, &l \le |k_z| + 1,\\
    (-1)^l,& l > |k_z| + 1.
  \end{cases}
\end{gather}
Note that (\ref{eq:trapped_area_and_wrapping_number})
implies that $c - (|k_z| + 1)$ is nonnegative and even.\\

{\it Anticonformal configurations} are given by antianalytic functions,
in analogy with the conformal case.  Specifically, if $f(w)$ is a
conformal configuration, then $\overline{ f(w)}$ is anticonformal with
invariants $(\ebar, \kbar, \Omegabar)$ given by
\begin{equation}
  \label{eq:barred_invariants}
  \ebar = (e_x, -e_y, e_z), \quad \kbar = (-k_x,
k_y, -k_z), \quad \Omegabar = -\Omega.
\end{equation}
Thus, conformal and anticonformal configurations are in one-to-one
correspondence.
Also, given any $(e,k,\Omega)$ and $(\ebar,\kbar,\Omegabar)$ related
as in (\ref{eq:barred_invariants}), 
one can verify from (\ref{eq:Omega_+}) and (\ref{eq:Omega_-})
that $\Omega_+(\ebar,\kbar) = \Omega_-(e,k)$, so that $\Omegabar \ge 
\Omega_+(\ebar,\kbar)$ if and only if $\Omega \le -\Omega_-(e,k)$.
Thus, conformal and anticonformal topologies are in one-to-one
correspondence, and representatives of every
anticonformal topology may be obtained from complex conjugation of the
associated conformal representative. 

The preceding discussion may be summarised as follows:
\begin{prop}
\label{prop:top and rep}
A reflection-symmetric topology  is conformal if and only if it
contains a  conformal configuration, and
anticonformal if and only if it contains a 
anticonformal configuration.
\end{prop}

\section{Upper bound for elastic energy}\label{sec: upper bound}
In \cite{mrz2004b} we showed that for a conformal or anticonformal
topology $(e,k,\Omega)$, 
\begin{equation}
  \label{eq:anti-con-bounds}
\Emin(e,k,\Omega) \le 8 L |\Omega|.
\end{equation}
where 
\begin{equation}
  \label{eq:L}
  L = (L_x^2 + L_y^2 + L_z^2)^{1/2}.
\end{equation}
We
note that the upper bound differs from the lower bound 
(\ref{eq:old_bound})
by a factor, $L/L_z$, which depends only on the
aspect ratios of the prism, and not on $(e,k,\Omega)$.

Here we derive an analogous upper bound for  nonconformal topologies.
To this end, we construct representatives $\nvec$. 
As in the conformal and anticonformal cases, we take these to
be reflection-symmetric and  radially constant in $\octant$.
In $\octant$,  $\nvec$ is taken to be of the form
\begin{equation}\label{eq:sterio-non}
  \left(\frac{n_x + i n_y}{1 + n_z}\right)(\rvec) = F(w,\wbar),\
  \text{where}\ w =
  \frac{x + iy}{r + z}.
\end{equation}
We take $F$ to be a juxtaposition of analytic and antianalytic domains
(in which the local estimate (\ref{eq:grad n^2}) for the energy
density becomes an equality), separated by an interpolating domain of
small energy.  
For definiteness, we take $\Omega < 0$ (the case $\Omega > 0$ is
treated analogously).  Let $f$ denote the
conformal configuration with topology $(e,k,-\Omega_-(e,k))$, so that
$f$ is the conformal configuration with the largest trapped area compatible
with $e$ and $k$.  Let
$w_0$ denote a point in the interior of $Q$, and let $D_\epsilon(w_0)
= \{w \,|\, |w - w_0| < \epsilon\}$ denote the open $\epsilon$-disk
about $w_0$.  Choose $w_0$ and $\epsilon$ so that $D_{2\epsilon}(w_0)$
is contained in $Q$ and contains no poles of $f$.  Let
\begin{equation}
  \label{eq:w_pm}
  W  = \fpi (\Omega + \Omega_-(e,k)).
\end{equation}
Since $(e,k,\Omega)$ is nonconformal, $W$ is a positive integer.  We
let
\begin{equation}
  \label{eq:F}
  F(w,\wbar) =
  \begin{cases}
    f(w), & w \in Q - D_{2\epsilon}(w_0),\\
 s 
f(w) + (1-
s)
(f(w_0) + (w-w_0)^W),&
    w \in D_{2\epsilon}(w_0) -  D_{\epsilon}(w_0),\\
    f(w_0) + \epsilon^{2W}(\wbar-\wbar_0)^{-W},&w \in
    D_{\epsilon}(w_0),
  \end{cases}
\end{equation}
where
\begin{equation}
  \label{eq:s}
  s(w,\wbar) = \frac{|w-w_0| - \epsilon}{\epsilon}
\end{equation}
(so that $s$ varies between $0$ and $1$ as $|w - w_0|$ varies between
$\epsilon$ and $2\epsilon$).  Thus, in $Q - D_{2\epsilon}(w_0)$, $F$
coincides with $f$ and therefore is conformal, while in
$D_\epsilon(w_0)$ it is anticonformal. 


Let us verify that $F$ has the required topology.
Since it coincides with $f$ on the boundary of $Q$, $F$ has the
same edge signs and kink numbers as $f$, namely $e$ and $k$. 
As for the trapped area, from (\ref{eq:trapped_area}) and
(\ref{eq:sterio-non}) it is straightforward to derive the general expression
\begin{equation}
  \label{eq:Omega w}
\Omega(F) = 
\int_Q 4 \frac{|\partial_\wbar F|^2 - |\partial_w F|^2
  }{(1 + |F|^2)^2}\, d^2 w.
\end{equation}
(\ref{eq:Omega w}) can be evaluated by dividing the domain of integration
as in (\ref{eq:F}).  The contribution from $Q -
D_{2\epsilon}(w_0)$ is, to $O(\epsilon^2)$, just the trapped area of
$f$, namely $-\Omega_-(e,k)$ (substituting $f$ for $F$ in (\ref{eq:Omega
  w}), the contribution from
$D_{2\epsilon}(w_0)$ is
$O(\epsilon^2)$).  Consider next the contribution from the disk
$D_\epsilon(w_0)$.  Here $F = f(w_0) + \epsilon^{2W}
(\wbar - \wbar_0)^{-W}$, so that $F$ covers the extended complex plane, 
apart from an $\epsilon^W$-disk about $f(w_0)$, $W$ times
with positive orientation.  It follows
that the contribution to (\ref{eq:Omega w}) is, to within 
 $O(\epsilon^W)$ corrections, $4\pi W$.
The remaining contribution, from the annulus 
$D_{2\epsilon}(w_0) - D_{\epsilon}(w_0)$, is $O(\epsilon^2)$.  This is
because the area of the annulus is $O(\epsilon^2)$, while the
integrand in (\ref{eq:Omega w}) may be bounded independently of
$\epsilon$ (by assumption, $f$ has no poles 
in $D_{2\epsilon}(w_0)$).  Since the trapped area is an odd multiple
of $\pi/2$, it follows that, for small enough $\epsilon$, 
\begin{equation}
  \label{eq:Omega(F)}
  \Omega(F) =
-\Omega_-(e,k) + 4\pi W = \Omega.
\end{equation}

By estimating the energy of the nonconformal representatives we can
obtain the following upper bound for $\Emin(e,k,\Omega)$:
\begin{thm}\label{thm: upper bound}
  Let $(e,k,\Omega)$ denote a nonconformal topology.  Then
  \begin{equation}
    \label{eq:upper bound nonconformal}
 \Emin(e,k,\Omega) \le 36  \pi L \sum_{\sigma} |w_\sigma|.
  \end{equation}
\end{thm}

\begin{proof}
From (\ref{eq:one-constant}) and (\ref{eq:sterio-non}) one can derive
the general expression for the energy,
  \begin{equation}
    \label{eq:E in w}
    E(F)
= 16\int_Q 4|\rvec(w)| \frac{ |\partial_\wbar F|^2 + |\partial_w F|^2 }{(1 + |F|^2)^2} \,
    d^2 w,
  \end{equation}
where $\rvec(w)$ is the point on the boundary of $\octant$ which
has $w$ as its stereographic projection.
Since $|\rvec(w)| \le L/2$,
it follows that
\begin{equation}
  \label{eq:E<..}
  E(F) \le 8 L \Acal(F),
\end{equation}
where
\begin{equation}
  \label{eq:Acal}
  \Acal(F) = \int_Q 
4\frac{ |\partial_\wbar F|^2 + |\partial_w F|^2 }{(1 + |F|^2)^2} \,
    d^2 w.
\end{equation}
$\Acal(F)$ represents the unoriented area of $\nvec(C_r)$.  The
expression (\ref{eq:Acal}) for $\Acal(F)$ differs from the expression
(\ref{eq:Omega w}) for $\Omega(F)$ in the relative sign of the $w$-
and $\wbar$-derivative terms.  (Thus, for conformal and anticonformal
configurations, one obtains the estimate (\ref{eq:anti-con-bounds}).)
Arguing as for (\ref{eq:Omega(F)}), we have that
\begin{equation}
  \label{eq:Acal_estimate}
  \Acal(F) \le |\Omega_-(e,k)|  + 4\pi W.
\end{equation}
From (\ref{eq:sum_of_ws}), this may be written as
\begin{equation}
  \label{eq:Acal_estimate_2}
  \Acal(F) \le 4\pi W -\half \pi \sum_\sigma w_{\sigma-},
\end{equation}
where $w_{\sigma-}$ are the (nonpositive) wrapping numbers of
$f$.
From
(\ref{eq:trapped_area_and_wrapping_number}) and (\ref{eq:w_pm}), the
$w_{\sigma-}$'s 
are related to the wrapping numbers of $F$, denoted $w_\sigma$,
according to
\begin{equation}
  \label{eq:wF_and_wf}
   w_{\sigma-} = w_{\sigma} - W.
\end{equation}
Substituting 
into (\ref{eq:Acal_estimate_2}), we
get that
\begin{equation}
  \label{eq:Acal_estimate3}
  \Acal(F)  \le  4\pi W + \half \pi  \sum_{\sigma} (W - w_\sigma) 
 \le  8 \pi W +  \half \pi \sum_\sigma |w_\sigma|.
\end{equation}
One easily establishes the estimate  
\begin{equation}
  \label{eq:W_estimate}
  2W \le \sum_\sigma |w_\sigma|.
\end{equation}
Indeed, since $\Omega < 0$ by assumption, it follows from 
(\ref{eq:sum_of_ws}) that 
\begin{equation}
  \label{eq:sigm_sum_1}
  \sum_{\sigma}w_\sigma < 0.
\end{equation}
Since $f$ has trapped area $-\Omega_-(e,k)$, it follows from 
Proposition~\ref{prop: 2.1}
that there is at least one octant, say $\sigma_0$, in which $f$ has zero
wrapping number.  From
(\ref{eq:wF_and_wf}), $w_{\sigma_0} = W$.  Let $\sum'_\sigma$ denote
the sum over octants with $\sigma_0$ omitted.  Then
\begin{equation}
  \label{eq:sigma_sum_2}
  {\sum_{\sigma}}' w_\sigma < -W.
\end{equation}
It follows that
\begin{equation}
  \label{eq:sigma_sum_3}
  \sum_\sigma |w_\sigma| =
{\sum_\sigma}' |w_\sigma| + W \ge \left|{\sum_\sigma}' w_\sigma\right| + W \ge 2W.
\end{equation}
Substituting (\ref{eq:sigma_sum_3}) into (\ref{eq:Acal_estimate_2}), we
get that
\begin{equation}
  \label{eq:Acal_3}
  \Acal(F) \le 9 \times  \half \pi \sum_\sigma |w_\sigma|.
\end{equation}
The required result, (\ref{eq:upper bound nonconformal}), follows from substitution into
 (\ref{eq:E<..}).
\end{proof}

\section{Discussion}\label{sec: discussion}
For nonconformal topologies, the ratio of the upper and lower bounds
for $\Emin$, as given by (\ref{eq:upper bound nonconformal}) and
(\ref{eq:infimum_result}), is $9 L/L_z$.  By finding representatives
of lower energy, it might be possible to obtain a ratio closer to the
conformal/anticonformal result, $L/L_z$ (which can be further improved
by more accurate estimates of the energy of the representatives \cite{mrz2004b}).

In \cite{mrz2004a}, we described, for conformal and anticonformal
topologies, a transition in topologically nontrivial equilibrium
(infimum-energy) configurations, from singular, in the case of a cubic
domain, to smooth, as the prism aspect ratios are varied.  Singular
configurations, when they appear, are limits of configurations which
differ from the topologically simplest ``unwrapped'' configurations
in thin tubes along the prism edges.  It would be interesting to
investigate whether similar transitions occur for nonconformal
topologies.  The nonconformal representatives differ from
conformal/anticonformal configurations only in a tube (a disk in the
two-dimensional stereographic description (\ref{eq:F})), and depending
on the aspect ratios, it may be energetically advantageous for these
tubes to collapse to edge singularities, or not.

AM was supported by an EPSRC/Hewlett-Packard Industrial CASE
Studentship.  MZ was partially supported by a grant from the Nuffield
Foundation.  We thank CJ Newton and A Geisow for stimulating our
interest in this area.

\appendix
\section{Kink numbers of conformal configurations}
\label{sec:appendix}
Taking $\nvec$ to be a conformal configuration
with stereographic projection $f$ given by
(\ref{eq:w(z)}), we derive formulas for the kink
numbers $k = (k_x,k_y,k_z)$ in terms of the parameters of $f$.\\

\noindent {\it Formulas for $k_x$ and $k_y$}.  For definiteness,
consider the calculation
of $k_y$.
Let $\rvec(\tau)$, $0 \le \tau \le 1$, denote a small quarter-circular arc
on the $xz$-face of $R$ starting on the $z$-axis and ending on the
$x$-axis (so that $\rvec(\tau)$ is positively oriented with respect to
the outward normal $-\yhat$ through the centre of the face).  Let $\nuvec(\tau) =
\nvec(\rvec(\tau))$.  Then $\nuvec(\tau)$ describes a curve on $S^2$
along the great circle in the $xz$-plane, starting from $e_z \zhat$
and ending at $e_x \xhat$.  $k_y$ is the winding number of
$\nuvec(\tau)$ relative to the shortest arc joining $e_z \zhat$ to
$e_x \xhat$, with anticlockwise windings about $-\yhat$ taken as positive.
$k_y$ is given by the number of times
$\nuvec(\tau)$ crosses a given point, say $\zhat$, counted with
a sign according to orientation.
Let $\tau_p$ denote the parameter values at these crossings.
Assuming that $\nuvec'(\tau_p)\ne 0$, we get that
\begin{equation}
  \label{eq:k_y-1}
  k_y = -\sum_{\tau_p> 0} \sgn(\nuvec'(\tau_p)\cdot\xhat)
+\half (1 + e_z)\cdot
\half (e_x - \sgn(\nuvec'(0)\cdot\xhat)).
\end{equation}
If $e_z = 1$ then $\nuvec(0) = \zhat$; the last term in
(\ref{eq:k_y-1}) accounts for a possible contribution in this case.
There is no contribution if $\sgn ( \nuvec'(0)\cdot \xhat) = e_x$, as
for the shortest arc joining $\zhat$ to $e_x \xhat$
(for which $k_y$ = 0).  Otherwise, the initial point constitutes a crossing
with sign $e_x$.  

Under stereographic projection, $\rvec(\tau)$ corresponds to 
$w(\tau) = \tau$,  the crossing $\nuvec(\tau_p) = \zhat$ corresponds to
$f(\tau_p) = 0$, 
and 
$\sgn (\nuvec'(\tau_p)\cdot\xhat)$ corresponds to $\sgn f'(\tau_p)$.
The zeros of $f$ on the real interval $(0,1)$ are given by the $r_j$'s
with $\rho_j = 1$.
Thus, (\ref{eq:k_y-1}) becomes
\begin{equation}
  \label{eq:k_y-2}
  k_y = -\sum_{j \,|\, \rho_j = 1} \sgn f'(r_j)
+ 
\half (1 + e_z)\cdot
\half(e_x - \sgn f'(0)).
\end{equation}
For simplicity, let us 
assume that 
the $r_j$'s are all distinct and ordered so that
$0 < r_1 <
\cdots < r_a < 1$.  
In this case, for $\rho_j = 1$, $f'(r_j) \ne 0$, and
from (\ref{eq:w(z)}), we have that
\begin{equation}
  \label{eq:f'(r_j)}
  \sgn  f'(r_j) = \sgn \left[ \epsilon  \frac{2r_j}{r_j^4 - 1} 
\prod_{m = 1\atop  m\ne j}^a  \left(\frac{r_j^2 - r_m^2}{r_m^2 r_j^2 - 1}
  \right)^{\rho_m}\right] = \epsilon(-1)^{j}.
\end{equation}
If $e_z = 1$,  $n$ is positive, and $w = 0$ is also a zero of $f$.  
As $f'(0)$ vanishes if $n > 1$, we replace $\sgn f'(0)$ by 
$\lim_{w\rightarrow 0} \sgn  f(w)  =  \epsilon$.
 (\ref{eq:k_y-2}) becomes
\begin{equation}
  \label{eq:k_y-3}
  k_y = -\epsilon  \sum_{j = 1}^a  (-1)^{j}\cdot\half (1+\rho_j) +
  \half (1 + e_z) \cdot \half (e_x - \epsilon)
\end{equation}
Recalling that $e_x = \epsilon (-1)^a$, with
some further straightforward
manipulation we obtain
\begin{equation}
  \label{eq:ky-final}
k_y = -\half (-1)^a e_x \left( \sum_{j = 1}^a (-1)^j \rho_j + \half
  e_z (1- (-1)^a)\right),
\end{equation}
which is just the expression given in (\ref{eq:ks}).
In fact, (\ref{eq:ky-final}) holds even if some of the $r_j$'s
coincide.

The expression for $k_x$ is similarly derived, with $\rvec(\tau)$ taken
to be a quarter-circular arc on the $zy$-face of $R$ with projection
$w = i(1-\tau)$.  Details are omitted.\\

\noindent {\it Formula for $k_z$}. 
Let $\rvec(\tau)$, $0 \le \tau \le 1$, denote a small quarter-circular arc
on the $xy$-face of $R$ starting on the $x$-axis and ending on the
$y$-axis (so that $\rvec(\tau)$ is positively oriented with respect to
the outward normal $-\zhat$ through the centre of the face).  Under
stereographic projection, $\rvec(\tau)$ corresponds to $w(\tau) =
\exp(i\pi \tau/2)$, and $\nvec(\rvec(\tau))$ to $f(\exp(i\pi \tau/2))$.  
$k_z$ is
the winding number of $f(\exp(i\pi\tau/2))$ on the unit circle in the
complex plane relative to the shortest arc joining $e_x \xhat$ to $e_y
\yhat$.  Clockwise windings are taken as positive.  It follows  that
\begin{equation}
  \label{eq:k_z_1}
  k_z = -\tpi \left(\arg f(\exp(i\pi/2)) - \arg f(0)\right) +
 \fourth \,\sgn(e_x e_y),
\end{equation}
where $\arg f(\exp(i \pi\tau/2))$ is taken to be continuous in $\tau$
and the last term ensures the winding number is zero for the shortest
arc joining $e_x \xhat$ to $e_y \yhat$.  Referring to (\ref{eq:w(z)}),
each factor of the form $[(w^2 \pm p^2)/(p^2 w^2 \pm 1 )]^{\xi}$
contributes $\xi \pi$ to the change in $\arg f$ in (\ref{eq:k_z_1}),
while $z^n$ contributes $n\pi/2$.  Thus we get
\begin{equation}
  \label{eq:k_z_2}
  k_z = -\half \sum_{j=1}^a \rho_j- \half \sum_{k =1}^b \sigma_k - \sum_{l=1}^c \tau_l
  - \fourth(n - e_x e_y),
\end{equation}
as in (\ref{eq:ks}).

\bibliography{lc05}
\end{document}